\documentclass[]{aa}

\usepackage{graphicx}
\begin{document}
   \title{A new seismic analysis of Alpha Centauri}

   \author{A.~Thoul\thanks{Chercheur Qualifi\'e
        au Fonds National de la Recherche Scientifique, Belgium.}, R.~Scuflaire, A.~Noels, B.~Vatovez,
   M.~Briquet, M-A.~Dupret, J. Montalban}

   \offprints{Anne.Thoul@ulg.ac.be}

   \institute{Institute of Astrophysics and Geophysics,
              University of Li\`{e}ge,
              17, all\'{e}e du 6-Ao\^{u}t,
              Sart Tilman (B\^{a}t B5c),
              B-4000 Li\`{e}ge 1
              }

   \date{}

   \abstract{
   Models of $\alpha$ Cen A \& B have been computed using the masses determined
   by Pourbaix et al.~(\cite{Po2002}) and the data derived from the spectroscopic
   analysis of Neuforge and Magain~(\cite{Ne1997}). The seismological data
   obtained by Bouchy and Carrier~(\cite{Bo2001}, \cite{Bo2002}) do help improve
   our knowledge of the evolutionary status of the system. All the constraints 
   are satisfied with a model which gives an age of about 6 Gyr for the binary.

   }

   \authorrunning{A. Thoul et al.}

   \maketitle
%

\section{Introduction}

The binary system Alpha Centauri offers a unique opportunity to test our
knowledge of stellar physics in solar-type stars other than the Sun. On
the one hand, its proximity allows very good determinations not only of
its parallax but also of its orbital parameters so that the masses and the
luminosities of both components can be determined.  On the other hand,
effective temperatures and  chemical compositions can be obtained
through spectroscopic analyses.

A number of theoretical calibrations of this system have already been
carried out. The reader will find references to earlier works in the
publications of Guenther and Demarque (2000) and Morel et al. (2000).
Very recently,  a new and very precise analysis by Pourbaix et al.
(2002) has led to new and strong constraints on the masses. Alpha~Cen~A
thus becomes an excellent candidate for asteroseismological studies since
solar-like $p$-mode oscillations have recently been discovered from
ground-based analysis (Bouchy and Carrier 2001, Carrier et al. 2002,
Bouchy and Carrier 2002).

In a very recent calibration and seismic analysis, Th\'evenin et al.
(2002) have found that the asteroseismological constraints, namely the
large and small frequency spacings, required a slight decrease in the
mass assigned by Pourbaix et al. (2002) to $\alpha$~Cen~A. Their model reproduces
the large and small frequency spacings although the fit in
the frequencies themselves is not achieved.

Using an entirely new evolution code,which does not yet include diffusion
of chemical elements, we
find models for the $\alpha$~Cen binary system which agree not only with all
the non-asteroseismic constraints, including the masses and the
requirement of the same age for both components, but also with the
asteroseismic constraints.

\section{Observational constraints}

Using data from different sources, Pourbaix et al.~(\cite{Po2002}) have recently
improved the precision of the orbital parameters of $\alpha$~Cen. They have
adopted the parallax derived by S\"{o}derhjelm~(\cite{So1999}),
$\varpi=747.1\pm1.2$~mas, and give for the masses $M_A/M_{\sun}=1.105\pm0.0070$
and $M_B/M_{\sun}=0.934\pm0.0061$.

From their spectroscopic analysis of the $\alpha$~Centauri system, Neuforge and
Magain (\cite{Ne1997}) have determined the effective temperature, the gravity
and the metallicity of both components. The effective temperatures are given by
$T_{{\rm eff},A}=5830\pm30$K and $T_{{\rm eff},B}=5255\pm50$K. The gravities are
given by $\log g_A=4.34\pm0.05$ and $\log g_B=4.51\pm0.08$.
Since the luminosity can be expressed in terms of the effective temperature and 
of the gravity through the relation
$$\log L/L_{\sun}=\log M/M_{\sun}+4\log T_{\rm eff}/T_{{\rm eff},{\sun}}
-\log g/g_{\sun}\,,$$
the error box in the $(\log T_{\rm eff},\log g)$ plane appears as a
parallelogram in the HR~diagram.

For the metallicity, they obtained
[Fe/H]$_A=0.25\pm0.02$ and [Fe/H]$_B=0.24\pm0.03$. To deduce $Z/X$, we assume
that this ratio is proportional to the abundance ratio
Fe/H and we adopt the solar value
$(Z/X)_{\sun}=0.023$ with an uncertainty of 10~\%, given by Grevesse and
Sauval~(\cite{Gr1998}). We then have the same value for both components
of $\alpha$~Cen: $Z/X=0.040\pm0.005$.
However, Grevesse and Sauval~(\cite{Gr2002}) now favour a
lower value $(Z/X)_{\sun}=0.0209$ with the same uncertainty. With this solar
value, we would have $Z/X=0.037\pm0.004$.  It seems safe to say that $Z/X$ is
between 0.033 and 0.045.

The above non-asteroseismic constraints are summarized in Table~\ref{ObsData}.

\begin{table}[h]
\centering
\caption{Non-asteroseismic constraints}
\label{ObsData}
\begin{tabular}{lcc}
\hline
                   &      A            &        B \\
\hline
$M/M_{\sun}$       & $1.105\pm0.0070$  & $0.934\pm0.0061$ \\
$\log T_{\rm eff}$ & $3.7657\pm0.0022$ & $3.7206\pm0.0041$ \\
$\log g$           & $4.34\pm0.05$     & $4.51\pm0.08$ \\
$Z/X$              &\multicolumn{2}{c}{$0.039\pm0.006$} \\
\hline
\end{tabular}
\end{table}

\section{The models}

A number of evolutionary sequences have been computed from the main sequence
with CL\'ES (Code Li\'egeois d'\'Evolution Stellaire).  We use the CEFF
equation of state (Christensen-Dalsgaard and D\"appen~\cite{Ch1992}).  The
opacities are those of the Lawrence Livermore National Laboratory (Iglesias and
Rogers ~\cite{Ig1996}) completed with the Alexander and Ferguson~(\cite{Al1994})
opacities at low temperature, both tables being smoothly joined in the 
temperature range defined by $3.95 < \log T <4.10$.  The nuclear reaction
rates are from Caughlan and Fowler~(\cite{Ca1988}) and the screening factor comes from
Salpeter~(\cite{Sa1954}).  The boundary conditions at the photosphere are
deduced from Kurucz (\cite{Ku1998}) models.  We use the mixing length theory of
convection. The convection parameter $\alpha=\ell/H_P$ is left free and is used to
force the evolutionary track through the error box around the
current position of the star in the HR~diagram. The suitable value of $\alpha$
depends on the initial chemical composition. It turns out that, for a given
composition, we can always use the same value of $\alpha$ for both components,
so that the equality $\alpha_A=\alpha_B$ comes out naturally and is not to be
considered as a supplementary constraint. The code does not include diffusion. 

\begin{figure}[h]
\centering
\resizebox{\hsize}{!}{\includegraphics{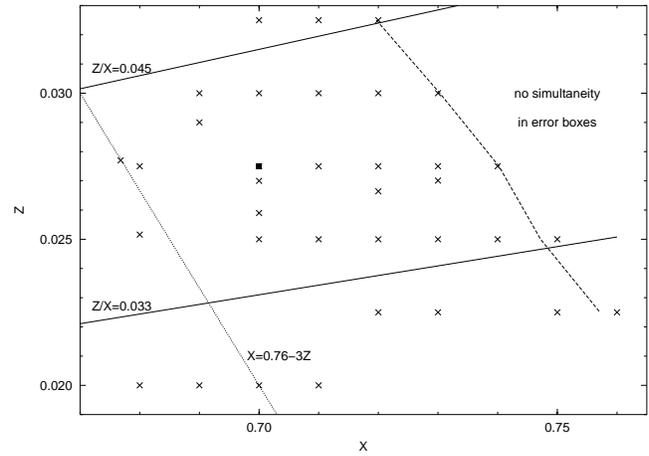}}
\caption{Constraints on the initial chemical composition
coming from the spectroscopy (solid line) and from the simultaneity condition
(dashed line). 
The dotted line represents the chemical evolution of the Galaxy.
The crosses indicate the different compositions for which evolution sequences have
been computed. The black square represents the model discussed in detail in this
paper.}
\label{chem}
\end{figure}

We have computed a number of models for both components, with different initial
chemical compositions, defined by $(X,Z)$, and convection parameter $\alpha$.
We require that both components of the binary system reach their respective
positions in the HR-diagram at the same age.  This requirement of simultaneity
determines a line in the $(X,Z)$-diagram to the right of which the B component
reaches its observed position in the HR-diagram after the A component has already 
left its own observed position (see Fig.~\ref{chem}).  
In addition, the chemical evolution of the Galaxy suggests a correlation
between $X$ and $Z$.  The relation $X=0.76-3Z$ which has been used by Pols et
al.~(\cite{Po1998}) is also shown in Fig.~\ref{chem}.  Although such a relation may
have a statistical significance, it may be hazardous to strictly adhere to it in
a particular case.
These constraints, together with the constraint on the $Z/X$ value, delimit a
permitted area in the $(X,Z)$ diagram.

\section{Asteroseismology of $\alpha$-Centauri}

Bouchy and Carrier~(\cite{Bo2001}), Carrier et
al.~(\cite{Ca2002}), and Bouchy and Carrier~(\cite{Bo2002})
have recently observed the oscillation spectrum
of $\alpha$~Cen~A with the CORALIE spectrograph. 
In the first paper, they observed the star 
during 5 nights, and were able to identify some $l=0$ and $l=1$ modes,
but the resolution was too low to identify separately the $l=0$
and the $l=2$ modes. In their next run of observations, they observed
the star during 11 nights, and were able to identify many $l=0$,
$l=1$ and $l=2$ modes. They could not resolve the $l=3$ modes.
Their latest results are summarized in Table~\ref{frequencies}.
They obtain an uncertainty of 0.46$\mu$Hz on these frequencies,
but add that an error of $\pm$1.3$\mu$Hz could have been introduced
for some of the frequencies due to aliases. For this reason, we will adopt 
this last value for the observational error bars.

\begin{table}
\caption{Mode frequencies (in $\mu$Hz) of $\alpha$~Cen~A.} 
\label{frequencies}
\begin{center}
\begin{tabular}{l|ccc|ccc}
\hline
 & \multicolumn{3}{c}{Observations$^a$} & \multicolumn{3}{c} {Our model}\\
\hline
 & $l$ = 0 & $l$ = 1 & $l$ = 2  & $l$ = 0 & $l$ = 1 & $l$ = 2 \\
\hline
$n$ = 15 &         &         & 1833.1 & 1730.5  & 1778.5  & 1828.0 \\
$n$ = 16 & 1841.3  & 1887.4  & 1934.9 & 1834.9  & 1882.8  & 1932.6 \\
$n$ = 17 &         & 1991.7  & 2041.5 & 1939.3  & 1987.7  & 2038.1 \\
$n$ = 18 &         & 2095.6  & 2146.0 & 2044.4  & 2093.7  & 2144.7 \\
$n$ = 19 & 2152.9  & 2202.8  & 2251.4 & 2150.8  & 2200.4  & 2251.7 \\
$n$ = 20 & 2258.4  & 2309.1  & 2358.4 & 2257.5  & 2307.5  & 2358.7 \\
$n$ = 21 & 2364.2  & 2414.3  & 2464.1 & 2364.2  & 2414.4  & 2465.9 \\
$n$ = 22 & 2470.0  & 2519.3  & 2568.5 & 2471.0  & 2521.5  & 2573.1 \\
$n$ = 23 & 2573.1  & 2625.6  &        & 2578.0  & 2629.0  & 2681.0 \\
$n$ = 24 & 2679.8  & 2733.2  & 2782.9 & 2685.5  & 2736.9  & 2789.2 \\
$n$ = 25 & 2786.2  & 2837.6  & 2887.7 & 2793.4  & 2845.2  & 2897.6 \\
\hline
\end{tabular}\\
\end{center}
$^a$ Bouchy and Carrier~(\cite{Bo2002}).
\end{table}

We have calculated the oscillation frequencies of $\alpha$~Cen~A using a
standard adiabatic code (Boury and al.~\cite{Bo1975}),
for a grid of models where the mass $M_A$ is fixed
at the value determined by Pourbaix et al.~(\cite{Po2002}),
i.e., $M_A=1.105 M_{\sun}$. The only free parameters
are $\alpha$, X, and Z, and the last two can only vary within the permitted area
illustrated
in Fig.~\ref{chem}.
The models we have explored are shown as crosses in Fig.~\ref{chem}. The models
which fall outside of the permitted area were explored for reason of
completeness only.

We calculate the modes $l=0$,
$l=1$ and $l=2$ for $n=15$ to 25. We then calculate the large frequency
spacings $\Delta\nu_{l,n}=\nu_{l,n}-\nu_{l,n-1}$ and the small frequency
spacings $\delta\nu_{0,n}=\nu_{l,n}-\nu_{l+2,n-1}$, and their averages
over $n=15$ to 25.

\section{Discussion and Conclusions}

\begin{figure}[h]
\centering
\resizebox{\hsize}{!}{\includegraphics{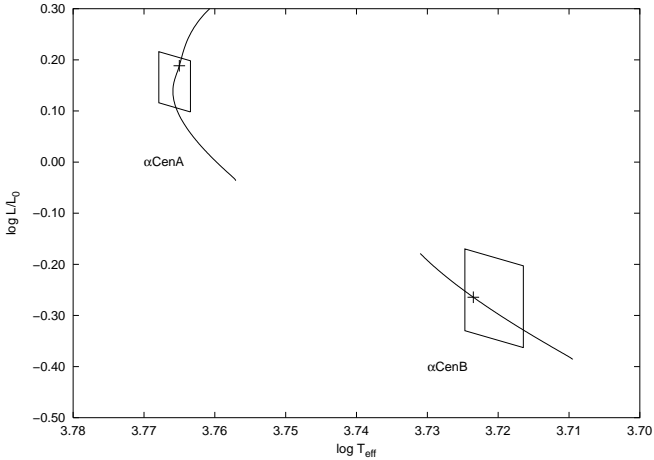}}
\caption{Evolutionary tracks in the HR diagram for the sequences
obtained with $X=0.70$, $Z=0.0275$, and $\alpha=1.8$.
The $1\ \sigma$ error boxes on $\log L/L_{\sun}$ and $\log T_{\rm eff}$ for
both components appear as parallelograms.
The crosses denote the model retained for our discussion.}
\label{hrdiagram49}
\end{figure}

\begin{figure}[h]
\centering
\resizebox{\hsize}{!}{\includegraphics{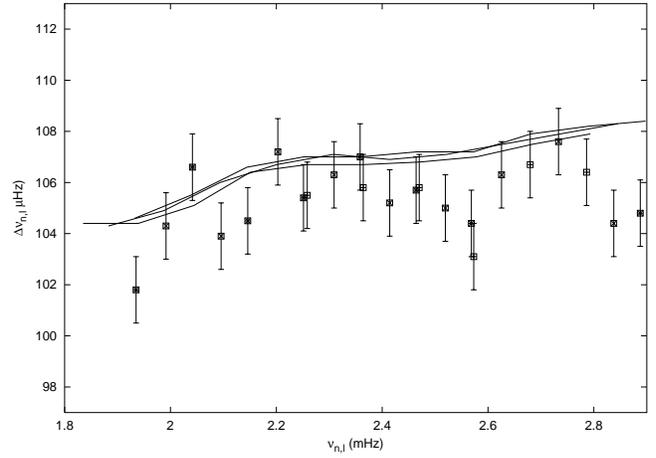}}
\caption{Large frequency spacing as a function of the frequency. The symbols
indicate the values obtained by Bouchy and Carrier (\cite{Bo2002}), with error bars of
$\pm 1.3\mu$Hz. The three lines correspond to the model shown as a cross in
Fig.(\ref{hrdiagram49}), for $l=0$, $l=1$ and $l=2$.}
\label{large49}
\end{figure}

\begin{figure}[h]
\centering
\resizebox{\hsize}{!}{\includegraphics{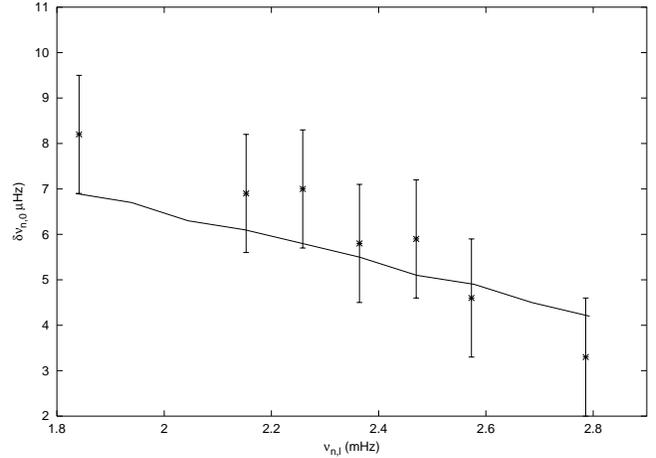}}
\caption{Small frequency spacing $\delta\nu_{0}$ as a function of the
frequency. The symbols represent the values obtained by Bouchy and
Carrier~(\cite{Bo2002}), with
error bars of $\pm 1.3\mu$Hz. The line corresponds to the model
shown as a cross in Fig.~\ref{hrdiagram49}.}
\label{small49}
\end{figure}

\begin{table}[h]
\centering
\caption{Our model}
\label{model}
\begin{tabular}{lcc}
\hline
                   &      A            &        B \\
\hline
$M/M_{\sun}$       & $1.105$           & $0.934$ \\
$L/L_{\sun}$       & $1.54$            & $0.544$ \\
$T_{\rm eff}$      & $5821$K           & $5291$K \\
$X$                &\multicolumn{2}{c}{$0.70$}   \\
$Z$                &\multicolumn{2}{c}{$0.0275$}   \\
$\alpha$           &\multicolumn{2}{c}{$1.8$}   \\
age                &\multicolumn{2}{c}{$6.41$Gyr}   \\
\hline
\end{tabular}
\end{table}

We will discuss the results corresponding to one of our best solutions 
(see Table~\ref{model}), shown as a
black square in Fig.~\ref{chem}.
Let us notice that for the same
physics, a solar calibration gives $\alpha=1.77$. 
The trajectories
in the HR diagram for the two components of the system are
shown in Fig.~\ref{hrdiagram49}.
The cross on the A component track
in the HR-diagram shows the model retained for our discussion.
We list the frequencies we have obtained for this particular
model in Table~\ref{frequencies}.
Finally, we show in Figures~\ref{large49} and \ref{small49} the results for
the large and the small frequency spacings. The age of this particular model is
$t=6.4$~Gyr. The position of the B component in the HR~diagram at the same age
is also marked with a cross in Fig.~\ref{hrdiagram49}.

This model satisfies all the non-asteroseismic
constraints listed in Table~\ref{ObsData}. Moreover,
the two components of the system have the same age.
As to the asteroseismic constraints, the average of the
large spacings is slightly larger than the one obtained from the observations,
but those large spacings remain within the error bars, as shown in Fig.~\ref{large49}.
The values for the small spacings fall within the error bars as well,
as shown in Fig.~\ref{small49}. Finally, we note that this solution
was found for a value of the mixing-length parameter $\alpha$ very close to the
solar value.

The value of the average small spacing $\langle\delta\nu_0\rangle$ is sensitive
to the structure of the core of the star, and therefore to its chemical composition. Let us note that our retained model does not have a convective core, but
it nevertheless leads to an excellent agreement for the small frequency spacings.
The value of $\langle\delta\nu_0\rangle$  increases as we
lower the value of X.
But all the models which lie on a curve more or less parallel
to the dashed curve in Fig.~\ref{chem}
provide the same value for $\langle\delta\nu_0\rangle$.
We are unable, given the resolution of the observations available, to
discriminate between those solutions.
To go one step further we would need additional observational data, such as the $l=3$ 
modes, or even better, the oscillation frequencies of $\alpha$~Cen~B.

We have been able with our model to reproduce quite well, and certainly within the
observational error bars,  both the large and the small frequency spacings observed by 
Bouchy et al. with the CORALIE spectrograph for $\alpha$ Cen A, while respecting
all the non-asteroseismic constraints of the system.  We must however be very careful 
with the interpretation of the results, since we have not included the diffusion of
helium and other heavy elements in our stellar evolution code.
It is well known that the inclusion of diffusion has greatly improved the agreement between
solar models and observations (see, e.g., Bahcall and Pinsonneault 1995, Morel et al. 1997). 
However, several papers have examined the influence of diffusion on models of $\alpha$ Cen A 
(Edmonds et al. 1992, Turcotte and Christensen-Dalsgaard 1998) and
they have shown that diffusion of helium has no detectable effect
on the frequencies of the low $l$-degree high-order  $p$-modes, 
and therefore no effect on the large and the
small frequency separations, {\it in the frequency range of the spectrum of $\alpha$ Cen A}. 
But it has a substantial effect on the age and on the initial 
composition. According to Edmonds et al. (1992) diffusion lowers the age of the star 
by about 10\%, and it increases the initial helium abundance to Y=0.30.

\begin{figure}[h]
\centering
\resizebox{\hsize}{!}{\includegraphics{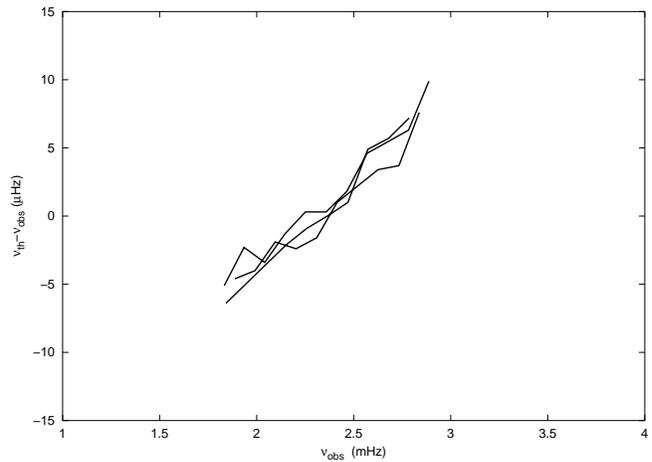}}
\caption{Differences between the calculated frequencies 
and the observed frequencies for our model}
\label{diff49}
\end{figure}

In their very recent paper, Th\'evenin et al.~(\cite{Th2002}) propose a model
for $\alpha$~Cen~A which gives acceptable values for the large and the small
frequency spacings, but they had to change the mass $M_A$ of the star.
In addition, their frequencies differ from the observed
frequencies by $\approx 30 \mu$Hz. 
It is well-known that the exact values of the frequencies themselves depend on the
details of the star's atmosphere, where the pulsation is non-adiabatic. A discrepancy of $30 \mu$Hz
seems rather large, however, since typical discrepancies for solar $p$-mode frequencies are 
smaller than $10 \mu$Hz (see, e.g., Guenther et al. 1996, Guenther \& Demarque 1997).
We show in Fig.~\ref{diff49} the differences between the calculated frequencies 
and the observed frequencies for our model. We see that these differences display a pattern
similar to what is found for the Sun (Guenther et al. 1996, Guenther \& Demarque 1997).

The evolution code used by Th\'evenin et al. (2002) differs
from ours in two ways: they use the Canuto \& Mazitelli (1991, 1992) theory of
convection (hereafter CM) below an optical depth $\tau=20$
while we use the basic B\"ohm-Vitense (1958) 
mixing-length theory of convection (hereafter MLT), 
and they include the effects of helium diffusion. 
Morel et al. (2000)
obtain very different ages for the $\alpha$ Cen system when using the 
CM or the MLT theories of convection: 4086 Myr and 2710 Myr respectively. 
In fact, changing the treatment of the convection, while keeping all other stellar
parameters the same, will affect the value of the effective temperature at a
given age (or luminosity). Therefore, in order to go through a given position in the
HR-diagram, it is necessary to modify, e.g., the helium abundance. Morel et al. (2000)
had to use Y=0.284 in MLT models, and Y=0.271 in CM models. This of course changes
considerably the age of the star. On the other hand, diffusion only changes 
the initial helium abundance by about 0.5\% (Turcotte and Christensen-Dalsgaard 1998), 
and the age by 0.4Gyr (Edmonds et al. 1992). Furthermore, the effects of microscopic 
diffusion could be even reduced if turbulent mixing is taken into account (Turcotte et al. 1998). 
In addition, Th\'evenin et al. (2002) use observations different from the ones we use 
to calibrate 
their models (Morel et al. 2000). 
The effective temperature they use for $\alpha$ Cen A is 
40K lower than ours. This will also affect the luminosity.
Moreover, they use a rather smaller value for the metallicity
[Fe/H]=0.20 instead of our [Fe/H]=0.25. The discrepancies between the results of
Th\'evenin et al. (2002) and the results we present here  could result from those differences.
It would of course be very useful to compare the results of different stellar evolution
codes in a more consistent and systematic manner.

\begin{acknowledgements}
\end{acknowledgements}
We thank J.~Christensen-Dalsgaard who kindly supplied us with the code for the
computation of the CEFF equation of state. This work has been supported by
the PRODEX-ESA/Contract\#15448/01/NL/SFe(IC).

\end{document}